\documentclass[10pt,twoside,BCOR7mm,DIV15,headinclude,footexclude,cleardoubleempty,idxtotoc]{scrartcl}

\usepackage[english]{babel}
\usepackage{graphicx}
\usepackage{hyperref}
\usepackage{scrpage2}
\usepackage{hyperref}
\usepackage{ifthen}

\makeatletter
\renewcommand{\@biblabel}[1]{}
\renewcommand{\@cite}[2]{%
{#1\ifthenelse{\boolean{@tempswa}}{,#2}{}}}
\makeatother

\hypersetup{breaklinks=true
,colorlinks=true,linkcolor=black,urlcolor=blue
,citecolor=black}

\ofoot{\thepage}
\ifoot{}

\setheadsepline{1pt}

\setkomafont{pagehead}{\normalfont\sffamily}
\setkomafont{pagenumber}{\normalfont\rmfamily}

\usepackage{booktabs}
\usepackage{amsmath}
\usepackage{amssymb}
\usepackage{multicol}
\usepackage{float}

\makeatletter
\newcommand{\listofcontributions}{\@starttoc{con}}

\newcommand{\l@contribution} {\@dottedtocline{1}{1.5em}{2.3em}}
\makeatother

\newenvironment{contribution}{
\setcounter{section}{0}
\setcounter{figure}{0}
\setcounter{table}{0}
\begin{flushleft}
{\em Clumping in Hot Star Winds \\
W.-R.\ Hamann, A.\ Feldmeier \& L.\ Oskinova, eds.\\
Potsdam: Univ.-Verl., 2007 \\
URN: http://nbn-resolving.de/urn:nbn:de:kobv:517-opus-13981
} 
\end{flushleft}
}{
\newpage
\lehead{}
\rohead{}
}

\begin{document}

\setlength{\baselineskip}{2.5ex}

\begin{contribution}


\lehead{M.\ Kraus, J.\ Kub\'at \& J.\ Krti\v{c}ka}

\rohead{Wind emission of OB supergiants and the influence of clumping}

\begin{center}
{\LARGE \bf Wind emission of OB supergiants \\ and the influence of clumping}\\
\medskip

{\it\bf M.\ Kraus$^1$, J.\ Kub\'at$^1$ \& J.\ Krti\v{c}ka$^2$}\\

{\it $^1$Astronomick\'y \'ustav, Akademie v\v{e}d \v{C}esk\'e republiky, Ond\v{r}ejov, Czech Republic}\\
{\it $^2$\'Ustav teoretick\'e fyziky a astrofyziky P\v{r}F MU, Brno, Czech Republic}

\begin{abstract}
The influence of the wind to the total continuum of
OB supergiants is discussed. For wind velocity distributions with $\beta >1.0$, 
the wind can have strong influence to the total continuum emission, even 
at optical wavelengths. Comparing the continuum emission of clumped and 
unclumped winds, especially for stars with high $\beta$ values, delivers flux
differences of up to 30\% with maximum in the near-IR. Continuum observations
at these wavelengths are therefore an ideal tool to discriminate between 
clumped and unclumped winds of OB supergiants.
\end{abstract}
\end{center}

\begin{multicols}{2}

\section{Introduction}

The spectra of hot stars show often excess emission at IR and radio 
wavelengths that can be ascribed to free-free and free-bound (ff-fb) emission 
from their wind zones (see e.g. Panagia \& Felli \cite{kraus:PanagiaFelli}).

Waters \& Lamers (\cite{kraus:WatersLamers}) have investigated this excess 
emission for $\lambda \ge 1\,\mu$m and winds with a $\beta$-law velocity 
distribution of varying $\beta$, pointing already to the sensitivity of 
the wind emission to the chosen velocity distribution.

Over the last few years, two major effects have become obvious that both
strongly influence the wind continuum emission: (i) the winds of hot stars
seem to be clumped, and (ii) many OB supergiants have winds with $1.0 \le \beta 
\le 3.5$ (see Table \ref{kraus:tab1}).

We investigate the wind continuum emission of OB supergiants especially at 
optical wavelengths. First, the influence of high $\beta$ values is discussed, 
and later on the effects of clumping are studied.

\section{Continuum of OB supergiants}

The calculation of the continuum emission of a typical OB supergiant is
performed in three steps: (i) we first calculate the stellar emission of the
supergiant with no stellar wind, (ii) then we calculate the emission of the
wind with the stellar parameters ($R_{*}, T_{\rm eff}$) as boundary conditions, 
(iii) and finally we combine the two continuum sources whereby the stellar 
emission still has to pass through the absorbing wind.
To simulate a typical OB supergiant we adopt the following set of stellar
parameters: $T_{\rm eff} = 28\,000$\,K, $R_{*} = 27.5\,R_{\odot}$, $\log
L_{*}/L_{\odot} = 5.62$, and $\log g = 3.1$. With these parameters we compute
the stellar continuum emission with the code of Kub\'at (\cite{kraus:Kubat}),
suitable for the calculation of NLTE spherically-symmetric model atmospheres.

\begin{table}[H]
\begin{center}
\caption{Range of $\beta$ values found for OB supergiants in the Galaxy (Markova et al. 
\cite{kraus:Markova} = Ma; Crowther et al. \cite{kraus:Crowther} = Cr; 
Kudritzki et al. \cite{kraus:Kudritzki} = Ku) and the Magellanic Clouds (Evans 
et al. \cite{kraus:Evans} = Ev; Trundle \& Lennon \cite{kraus:TrundleLennon} = 
TL; Trundle et al. \cite{kraus:Trundle} = Tr).}
\label{kraus:tab1}
\medskip
\begin{tabular}{lcl}
\toprule
Sp.\,Type    & $\beta$   & Ref. \\
\midrule
O4 -- O9.7 & 0.7 -- 1.25 & Ma \\
O9.5 -- B3 & 1.2 -- 3.0 & Cr \\
B0 -- B3 & 1.0 -- 3.0 & Ku \\
O8.5 -- B0.5 & 1.0 - 3.5 & Ev \\
B0.5 -- B2.5 & 1.0 - 3.0 & TL \\
B0.5 -- B5 & 1.0 - 3.0 & Tr \\
\bottomrule
\end{tabular}
\end{center}
\end{table}

The spherically symmetric wind is assumed to be fully ionized, isothermal, and
in LTE. This reduces the problem to a pure 1D treatment of the simplified
radiation transfer (e.g. Panagia \& Felli \cite{kraus:PanagiaFelli}).
The electron temperature is fixed at 20\,000\,K, and the density distribution
in the wind follows from the equation of mass continuity, relating the density
at any point in the wind to the mass loss rate and the wind velocity.

The velocity of hot star winds can be approximated with a $\beta$-law
\begin{equation}
v(r) = v_{0} + (v_{\infty} - v_{0}) \left( 1 - \frac{R_*}{r}\right)^{\beta}
\label{velo}
\end{equation}
where $\beta$ describes the steepness of the velocity increase at the base of
the wind, and $v_{0}$ defines the velocity on the stellar surface.
A more detailed description of the calculations will be given elsewhere.

\subsection{The influence of the velocity}
                                                                                
The range in $\beta$ found for galactic and Magellanic Cloud OB supergiants is 
listed in Table\,\ref{kraus:tab1}. Increasing $\beta$ means that the wind is
accelerated more slowly. Consequently, the density in these regions is enhanced
because $n_{\rm e}(r) \sim \dot{M}/v(r)$. These density peaks close to the
stellar surface with respect to the wind density with $\beta = 1.0$ are shown 
in Fig.\,\ref{kraus:dens}. 

Even though these density peaks are rather narrow in radius, they strongly
influence the optical depth and therefore the emission of the ff and 
especially the fb processes. Now, the wind can become (at least 
partially) optically thick even at optical wavelengths. This leads to an 
enhanced wind emission meanwhile the stellar flux suffers from the 
simultaneously increasing wind absorption.

Our test supergiant is assumed to have a wind with $\dot{M} = 5\times 
10^{-6}\,M_{\odot}$yr$^{-1}$, $v_{\infty} = 1550$\,km\,s$^{-1}$, and we 
calculate the continuum emission for $\beta = 1, 2$, and 3. The results are 
shown in Fig.\,\ref{kraus:beta}. It is obvious that with increasing $\beta$ 
the wind creates a near-IR excess, absorbs more of the stellar emission, and 
contributes to the total emission even at optical wavelengths.

\begin{figure}[H]
\begin{center}
\includegraphics[width=\columnwidth]{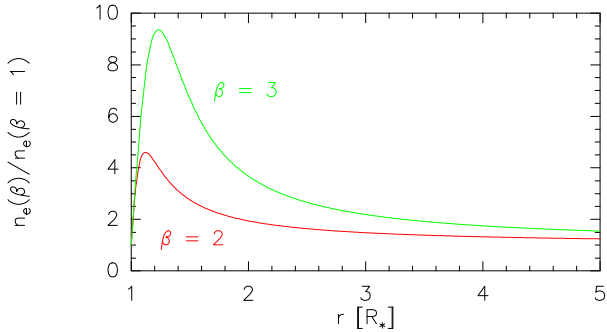}
\caption{Pronounced density peaks close to the surface (compared to the density 
for $\beta = 1$) that grow with increasing $\beta$.\label{kraus:dens}}
\end{center}
\end{figure}

\subsection{The influence of clumping}

Hillier et al. (\cite{kraus:Hillier}) introduced a formalism to account for
the presence of wind clumping, and in our calculations we use their 
filling factor defined by 
\begin{equation}
f(r) = f_{\infty} + (1 - f_{\infty}) e^{-v(r)/v_{\rm cl}}
\end{equation}
and setting $f_{\infty} = 0.1, v_{\rm cl} = 30$\,km\,s$^{-1}$, and $v(R_*)=
v_{\rm thermal}$.
Since $f$ depends on the velocity distribution, this filling factor 
is a function of radius as well, constructed such that it quickly reaches its
terminal value (top panel of Fig.\,\ref{kraus:dens_cl}). This way of clumping 
introduction requests, however, that in order to maintain the same radio flux,
the mass loss rate has to be decreased, i.e. $\dot{M}_{\rm cl} = 
\sqrt{f_{\infty}}~\dot{M}_{\rm smooth}$, while the absorption coefficient of 
the ff-fb processes increases, $<\kappa>_{\rm cl} = f(r)^{-1}~\kappa_{\rm 
smooth}$, due to its density squared dependence. Our clumped models 
automatically account for this mass loss reduction.

At those positions in the wind where $f(r) = f_{\infty}$ there is no difference
between the clumped and the unclumped wind. However, in those regions where
$f(r) \neq f_{\infty}$, which are also the regions where $\beta$ has its 
strongest influence, the wind opacity is sensitive to the clumping. But while
$\beta$ enhances the density, clumping (for the same input $\dot{M}_{\rm 
smooth}$) reduces the density again. A wind with 
high $\beta$ and clumping will therefore have a different opacity in the 
innermost wind region than a wind with low $\beta$ and clumping, and the 
clumped wind will have a different opacity than the smooth wind. This is
shown in the lower panel of Fig.\,\ref{kraus:dens_cl} where we plotted the 
opacity ratio of the clumped with respect to the smooth wind for different 
values of $\beta$. The higher $\beta$, the stronger the effect.
In Fig.\,\ref{kraus:clumped} we compare the continuum of a clumped 
with an unclumped wind for $\beta = 3.0$.
\begin{figure}[H]
\begin{center}
\includegraphics[width=\columnwidth]{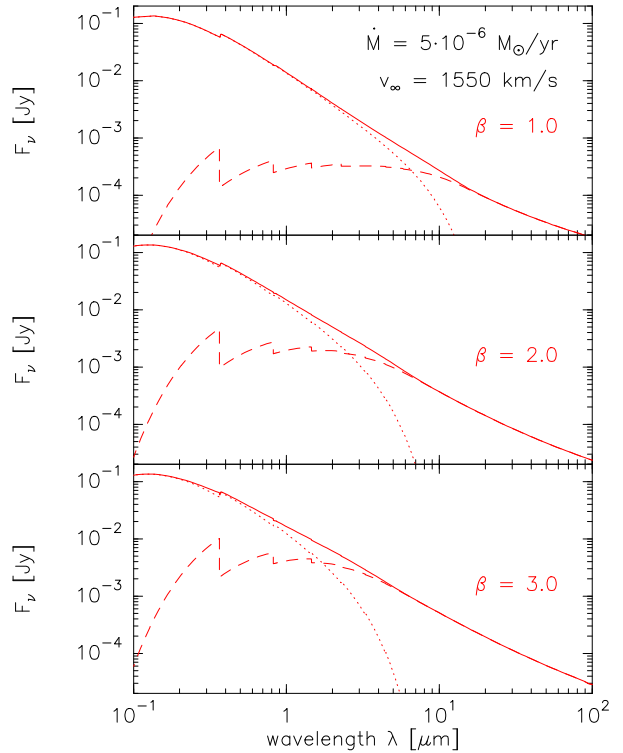}
\caption{Continuum emission of the test OB supergiant for different values
of $\beta$. Shown are the stellar emission having passed through the absorbing
wind (dotted), the ff-fb emission from the wind (dashed)
and the total continuum (solid).
\label{kraus:beta}}
\end{center}
\end{figure}
\begin{figure}[H]
\begin{center}
\includegraphics[width=\columnwidth]{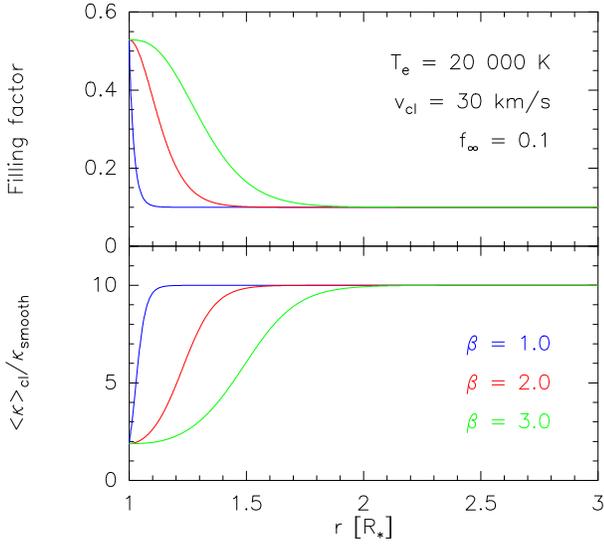}
\caption{Top: Filling factor for different values of $\beta$. Bottom:
Opacity ratio between clumped wind model and unclumped wind model. 
\label{kraus:dens_cl}} 
\end{center}
\end{figure}

It is obvious that the clumped model generates less wind emission for $\lambda 
< 10\,\mu$m. For a better visualization we calculated the flux ratios between
unclumped and clumped models (Fig.\,\ref{kraus:ratio}). They show a maximum of
up to 30\% at near-IR wavelengths. Continuum observations at these wavelengths
are therefore an ideal tool to discriminate whether the winds of OB stars
with $\beta > 1.0$ are clumped. 

\begin{figure}[H]
\begin{center}
\includegraphics[width=\columnwidth]{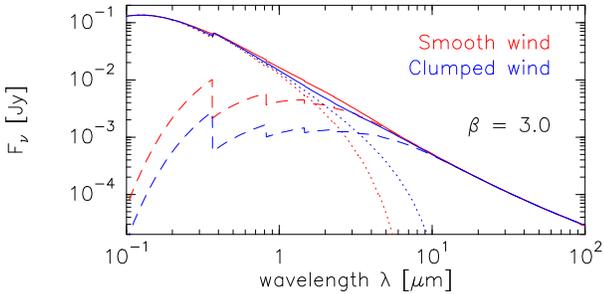}
\caption{Clumped versus unclumped wind with $\beta = 3.0$.
The clumped model produces less wind emission for $\lambda < 10\,\mu$m, 
resulting in a lower total near-IR flux.
\label{kraus:clumped}}
\end{center}
\end{figure}

\section{Conclusions}

For OB supergiants with high $\beta$ values the ff and especially the fb 
emission can strongly influence the total continuum, even at optical 
wavelengths.

Clumping, introduced by the filling factor approach, also influences the wind 
opacity and therefore the continuum emission. Whether the wind of an OB 
supergiant is clumped or not can be checked based on continuum 
observations in the optical and near-IR region. Especially winds with high 
$\beta$ are found to have fluxes that differ by about 30\% (see 
Fig.\,\ref{kraus:ratio}). The optical and near-IR continuum are therefore  
ideal to discriminate between clumped and unclumped winds.

\begin{figure}[H]
\begin{center}
\includegraphics[width=\columnwidth]{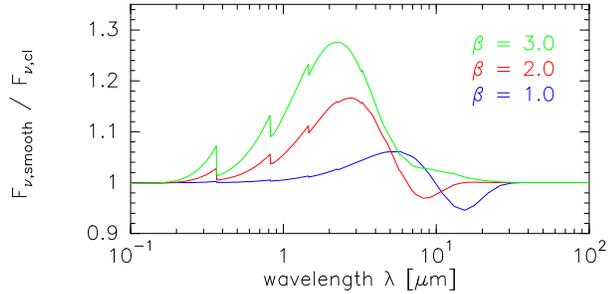}
\caption{Continuum flux ratio between the unclumped and clumped wind models.
The ratio increases with $\beta$ having a maximum in the near-IR. 
\label{kraus:ratio}}
\end{center}
\end{figure}

\paragraph{Acknowledgements}

M.K. acknowledges financial support from GA\,AV grant number KJB\,300030701.


\end{multicols}

\end{contribution}


\end{document}